\documentclass[reqno,12pt]{article}
\usepackage{ae} 
\usepackage[T1]{fontenc}
\usepackage[ansinew]{inputenc}
\usepackage{amsmath}
\usepackage{amssymb}
\usepackage{graphicx}
\usepackage{color}

\definecolor{darkblue}{cmyk}{0.9,0.9,0,0}
\usepackage[colorlinks=true,linkcolor=darkblue,citecolor=darkblue,urlcolor=darkblue]{hyperref}
\usepackage{epsfig}

\newcommand{\beq}{\begin{equation}}
\newcommand{\eeq}{\end{equation}}
\newcommand\beqa{\begin{eqnarray}}
\newcommand\eeqa{\end{eqnarray}}
\newcommand\bea{\begin{array}}
\newcommand\eea{\end{array}}

\newcommand{\bPhi}{\bar\Phi}
\newcommand{\im}{{\rm Im}\;}

\def\Xint#1{\mathchoice
{\XXint\displaystyle\textstyle{#1}}%
{\XXint\textstyle\scriptstyle{#1}}%
{\XXint\scriptstyle\scriptscriptstyle{#1}}%
{\XXint\scriptscriptstyle\scriptscriptstyle{#1}}%
\!\int}
\def\XXint#1#2#3{{\setbox0=\hbox{$#1{#2#3}{\int}$}
\vcenter{\hbox{$#2#3$}}\kern-.5\wd0}}

\def\dashint{\Xint-}

\newcommand{\neqa}{\nonumber\end{eqnarray}}
\newcommand{\la}[1]{\label{#1}}

\def\tr{{\rm tr~}}

\newcommand{\<}{{\langle}}
\renewcommand{\>}{{\rangle}}

\newcommand{\re}{\relax{\rm I\kern-.18em R}}

\def\su2{{SU(2)}}

\def\[{\left[}
\def\]{\right]}

\def\[{\left[}
\def\]{\right]}

\def\<{\langle}
\def\>{\rangle}

\def\i2{\frac{i}{2}}

\def\bPhi{{\bar \Phi}}

\def\bPhi{{\bar \Phi}}

\usepackage[font=small,labelfont=bf,width=0.85\textwidth]{caption}

\usepackage{varioref}
\usepackage{makeidx}
\makeindex

\usepackage[english]{babel}

\begin{document}


\thispagestyle{empty}

\renewcommand{\thefootnote}{\fnsymbol{footnote}}
\setcounter{footnote}{0}
\setcounter{figure}{0}
\begin{center}
{\Large\textbf{\mathversion{bold} 
 Unified approach to the $ SU(2) $ Principal Chiral Field model at Finite Volume
}\par}

\vspace{1.0cm}

\textrm{Jo\~ao Caetano}
\\ \vspace{1.2cm}
\footnotesize{
\textit{Perimeter Institute for Theoretical Physics\\ Waterloo,
Ontario N2J 2W9, Canada\\
\vspace{3mm}
Department of Physics and Astronomy \& Guelph-Waterloo Physics Institute,\\University of Waterloo, Waterloo, Ontario N2L 3G1, Canada\\
\vspace{3mm}
Centro de F\'\i sica do Porto e Departamento de F\'\i sica e Astronomia\\ Faculdade de Ci\^encias da Universidade do Porto, \\
Rua do Campo Alegre, 687, \,4169-007 Porto, Portugal} \\
\texttt{jd.caetano.s AT gmail.com}
\vspace{3mm}}


\par\vspace{1.5cm}

\textbf{Abstract}\vspace{2mm}
\end{center}

\noindent
\small
Typically, the exact ground state energy of integrable models at finite volume can be computed using two main methods: the thermodynamic Bethe ansatz approach and the lattice discretization technique. For quantum sigma models (with non-ultra local Poisson structures) the bridge between these two approaches has only been done through numerical methods. We briefly review these two techniques on the example of the $ SU(2) $ principal chiral field model and derive a single integral equation based on the Faddeev-Reshetikhin discretization of the model. We show that this integral equation is equivalent to the single integral equation of Gromov, Kazakov and Vieira derived from the TBA approach.

\vspace*{\fill}

\setcounter{page}{1}
\renewcommand{\thefootnote}{\arabic{footnote}}
\setcounter{footnote}{0}

\newpage

\section{Introduction}
The spectrum of quantum integrable models in 1+1 dimensions can be computed exactly. In an infinite space (line) the spectrum is continuous and the relevant quantities are the dispersion relation $\epsilon(p)$ of the fundamental particles and their two-body $S$-matrix $S(p,p')$.\footnote{We are oversimplifying since in general particles might have additional internal degrees of freedom and the $S$-matrix is a matrix. The $n$-body $S$-matrix factorizes for integrable models and therefore the main building block is the two-body $S$-matrix.} In a finite circle the spectrum of the system is quantized. When the circle length is very large, the momenta of the particles are constrained by a set of asymptotic Bethe equations
\beq
e^{ip_j L} \prod_{k\neq j}^N S(p_j,p_k)=1
\eeq
and the energy is given by the sum of the individual energies of the particles, $E=\sum_{j=1}^N \epsilon(p_j)$. This treatment is not exact in the system size. If $L$ is not very large, quantum corrections associated to virtual particles going around the space-time cylinder invalidate the asymptotic Bethe ansatz treatment. For relativistic systems, these corrections are typically or order $e^{-mL}$ where $m$ is the mass of the lightest particle \cite{Luscher:1985dn,Luscher:1986pf}. 

There are two main approaches to study the exact spectrum of quantum integrable theories at finite volume. One is based on a Wick rotation trick devised by Zamolodchikov \cite{Zamolodchikov:1989cf}. The idea is to exchange time and space in the path integral. The finite circle length leads to finite temperature after Wick rotation. The infinite time over which we preform the path integral becomes an infinite space extent. In the Wick rotated picture the asymptotic Bethe ansatz description is therefore exact. This trick can be made rigorous for the computation of the ground state. Under some assumptions, it can be generalized to excited states \cite{Dorey:1996re,Bazhanov:1994ft}. 

The second approach is based on integrable lattice discretizations of the quantum field theory. Frequently, we can interpret particles excitations of the integrable model as effective excitations over a non-trivial vacuum of a much simpler bare theory. Often, the bare theory is the continuum limit of a quantum spin chain. The effective particles of the field theory are the spinons of the quantum spin chain, i.e. the spin chain excitations around its anti-ferromagnetic vacuum. The bare theory is typically ultra local and admits an exact Bethe ansatz solution. The exact spectrum of the effective theory can be studied by carefully following the continuum limit of these exact equations.

For many examples, it was understood how to match these two approaches analytically. However a systematic understanding of the connection of the two methods is still lacking. In particular, for quantum sigma models with non-ultra local Poisson structures the bridge between these two approaches has only been done through numerical methods. It is the purpose of this paper to preform a first analytical comparison of the two methods. For illustration we will consider the example of the exact ground state energy of the $SU(2)$ principal chiral field \cite{Polyakov:1983tt,Polyakov:1984et}. 

The exact vacuum energy\footnote{The authors considered not only the vacuum energy but also all the excited states however as explained above we will focus on the ground state energy in this work.} $ E_0 $ of the $SU(2)$ principal chiral field was recently considered by Gromov, Kazakov and Vieira (GKV) through the Wick rotation approach. The outcome of this approach is a single integral equation given by 
\begin{equation}
\begin{gathered}
g^2=-e^{i m L \sinh(\pi u)} \exp \left(2i\; \im\! \[K_0^-*\log\frac{\left(g^+\right)^2-1}{ | g^+ |^2-1} \] \right)\;\label{eq:g2} \\
E_0=-\frac{m}{2}\int du\cosh(\pi u)\log\frac{\left(g^+\right)^2-1}{ | g^+ |^2-1}\frac{\left(\bar{g}^-\right)^2-1}{ | g^+ |^2-1}\,.
\end{gathered} 
\end{equation}
The star denotes convolution, the superscripts $\pm$ indicate shifts of the functions in the imaginary direction and $K_0$ is the Kernel of the $SU(2)$ $S$-matrix. These quantities are rigorously defined in the main sections of the text.  

The lattice discretization approach started with the seminal work of Faddeev and Reshetikhin \cite{Faddeev:1985qu}. In this work the authors proposed a discretization of the $SU(2)$ principal chiral field, whose diagonalization is equivalent to the one of an inhomogeneous  $ SU(2) $ spin s chain. This is quite an interesting proposal with several subtleties reviewed in section \ref{FR}. The computation of the exact spectrum of the $SU(2)$ principal chiral field can be done using this discretization. This was done in \cite{Hegedus:2003xd}\footnote{Actually, the approach in \cite{Hegedus:2003xd} is based on the light-cone discretization proposed in \cite{Destri:1987ug}, which turns out to be equivalent to the FR model \cite{Destri:1987ug,Destri:1987zu}} leading to a set of two coupled integral equations of the form\footnote{The convolutions with $ K_{0}^{++} $ appearing in these equations are understood in the principal part sense}
\begin{equation}
\begin{gathered}
\begin{aligned}
\log a_{0}=&-mL\cosh{(\pi u)}+\frac{1}{2}K_{0}\ast\log(1+a_{0})-\frac{1}{2}K_{0}^{++}\ast\log(1+\bar{a}_{0})\\
&-\frac{1}{2}\log(1+\bar{a}_{0})+s\ast\log(1+a)(1+\bar{a})\end{aligned}\\
\log a\,=\,\frac{1}{2}K_{0}\ast\log(1+a)-\frac{1}{2}K_{0}^{++}\ast\log(1+\bar{a})-\frac{1}{2}\log(1+\bar{a})+s\ast\log(1+a_{0})(1+\bar{a}_{0}) \\
E_0=-\frac{m}{2}\int du\cosh(\pi u)\log{(1+a_{0})(1+\bar{a}_{0})}\,.
\end{gathered} \la{intro}
\end{equation}
Numerical analysis indicates that the Wick rotation approach and the lattice discretization approach agree \cite{Gromov:2008gj,Beccaria:2010gq}. In this paper, we show that in fact this set of equations can be reduced to a single one. Moreover, we also show that this new single integral equation is equivalent to (\ref{eq:g2}), once a proper change of variable is performed.

We will also discuss what is the reason for the seemingly more complicated structure of (\ref{eq:g2}) compared to the usual Destri-de Vega equation \cite{Destri:1992qk,Destri:1994bv}.\footnote{More precisely, in the usual DdV approach we typically obtain (\ref{eq:g2}) without the denominator factors appearing in the logarithms of those equations.} As explained in the last section this difference is directly related to the structure of the spin chain vacuum which is given by a condensate of bound-states known as Bethe strings. In the cases where DdV holds, for example for the $SU(2)$ Gross-Neveau model, the vacuum distribution is typically given by a sea of real momenta particles. 

The structure of the paper is as follows: in section 2, we introduce the model and expose the necessary tools for the GKV method; the section 3 will be devoted to the origin of the lattice formulation of the model, and to the description of the procedure to obtain the corresponding integral equation; finally, in section 4, we show the equivalence of the two approaches and give some physical insight about the meaning of the obtained equations.


\section{Review of the Wick rotation approach}
\subsection{The $SU(2)$ Principal Chiral model}
The two dimensional $SU(2)$ Principal Chiral model is defined by the following action
\begin{equation}\label{eq:action}
S=-\frac{1}{2e_{0}^{2}}\int{dt\,dx\,\tr{(g^{-1}\partial_{\mu}{g})^{2}}}\,,
\end{equation} with $g\in SU(2)$. It has the global symmetry $ SU(2)_{L}\times SU(2)_{R}$ which corresponds to the left and right multiplications by constant matrices $g_{0},g_{1}\in SU(2)$, i.e. $g\rightarrow g_{0}\,g\,g_{1}$. We consider the space direction to be a circle of length $L$. 
In infinite volume, the fundamental excitations are massive relativistic particles transforming in the bifundamental representation of $ SU(2)\times SU(2)$. We use the rapidity $\theta$ to parametrize the energy and momentum of these excitations as 
\beq
\epsilon(\theta) \pm p(\theta)=m\, e^{\pm \pi \theta} \,.
\eeq
The two body $S$-matrix describing the scattering of these particles is completely fixed by Yang-Baxter, unitarity, crossing symmetry and absence of bound-states and is given by \cite{Zamolodchikov:1978xm} 
\begin{equation}
S(\theta_1,\theta_2)=\frac{S_{0}(\theta)}{\theta-i}(\theta\,\mathbb{I}+i\,\mathbb{P})\otimes\frac{S_{0}(\theta)}{\theta-i}(\theta\,\mathbb{I}+i\,\mathbb{P}) \qquad \text{where} \qquad S_{0}(\theta)=i\frac{\Gamma\left[\frac{1}{2}-\frac{i\theta}{2}\right]\Gamma\left[+\frac{i\theta}{2}\right]}{\Gamma\left[\frac{1}{2}+\frac{i\theta}{2}\right]\Gamma\left[-\frac{i\theta}{2}\right]}\,,
\end{equation}
with $\theta=\theta_1-\theta_2$. 

\begin{figure}
\centering
\scalebox{0.65}{  
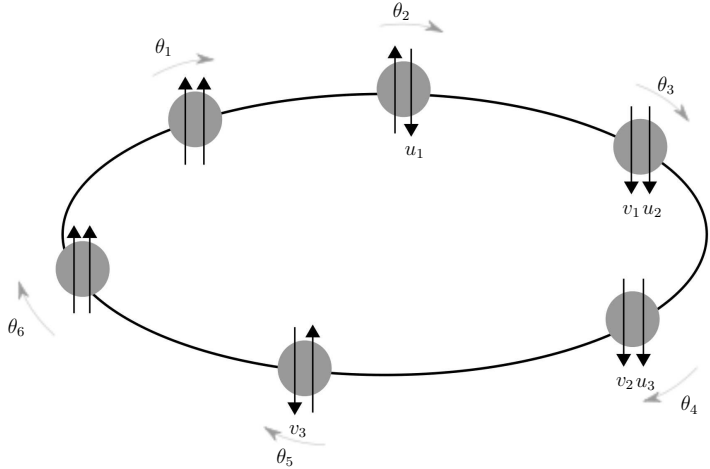}
\caption{Schematically, the $ SU(2) $ PCF describes massive particles with momenta parametrized by $ \theta $, and internal degrees of freedom corresponding to bifundamental representations of $ SU(2)\times SU(2) $. Exciting these isotopic sectors is equivalent in this picture to flipping the corresponding spins.}
\label{ring}
\end{figure}

Next we move to the large volume ($Lm\gg 1$) spectrum. We consider $N$ particles in the circle. The simplest possible configuration for the $SU(2)_L\times SU(2)_R$ internal degrees of freedom is to have all the left spins and all the right spins pointing up. A generic configuration is obtained from this one by flipping $N_L$ left spins and $N_R$ right spins. The large volume ($Lm\gg 1$) spectrum is obtained from the periodicity of the multi-particle wave function. This condition takes the form of a set of (nested) Bethe ansatz equations \cite{Gromov:2006dh} that entangle the several degrees of freedom of the particles: the physical rapidities ($ \theta's $) and the internal rapidities ($ u's $ and $ v's $) (see figure \ref{ring})
\begin{equation}\label{eq:ABA1}
e^{ip_{j}L}\prod_{k\neq j}^{N}S_{0}^{2}(\theta_{j}-\theta_{k})\prod_{k=1}^{N_{R}}\frac{\theta_{j}-u_{k}+\frac{i}{2}}{\theta_{j}-u_{k}-\frac{i}{2}}\prod_{k=1}^{N_{L}}\frac{\theta_{j}-v_{k}+\frac{i}{2}}{\theta_{j}-v_{k}-\frac{i}{2}}=1
\end{equation}
\begin{equation}\label{eq:ABA2}
\prod_{k=1}^{N}\frac{u_{j}-\theta_{k}-\frac{i}{2}}{u_{j}-\theta_{k}+\frac{i}{2}}=\prod_{k\neq j}^{N_{R}}\frac{u_{j}-u_{k}-i}{u_{j}-u_{k}+i}
\end{equation}
\begin{equation}\label{eq:ABA3}
\prod_{k=1}^{N}\frac{v_{j}-\theta_{k}-\frac{i}{2}}{v_{j}-\theta_{k}+\frac{i}{2}}=\prod_{k\neq j}^{N_{L}}\frac{v_{j}-v_{k}-i}{v_{j}-v_{k}+i}\,.
\end{equation}
The asymptotic spectrum is simply the sum of the free dispersion relation of the particles\begin{equation}
E=\sum_{k=1}^{N}m\cosh{(\pi \theta_{k})}\,.
\end{equation}
For smaller volumes the asymptotic Bethe ansatz description breaks down and new techniques need to be developed. The simplest quantity we can consider in finite volume is the ground-state energy of the theory, or Casimir energy. 
The most common procedure to compute this quantity is by means of the Thermodynamic Bethe Ansatz (TBA), which was originally developed to compute finite temperature thermodynamics of integrable models, i.e. its free energy. If we perform a double Wick rotation and simultaneously interchange the role of time and space variables we realize that the free energy per unit length at temperature $1/L$ actually coincides with the  ground-state energy at finite volume $L$ \cite{Zamolodchikov:1989cf}.\footnote{The double Wick rotation does not change the theory due to Lorentz invariance. For non-Lorentz invariant theories the Wick rotation trick can often still be applied but it would relate the free energy of one theory with the ground-state energy of another theory.} 

Once we consider the system at finite temperature and infinite length we excite an infinite number of all possible excitations, $N,N_L,N_R\to \infty$. In this limit, if some rapidity $ u_j $ has an imaginary part, the left hand side of (\ref{eq:ABA2}) diverges (or decays to zero). The right hand side must have the same behaviour. Hence, there ought to be another rapidity separated by $ i $ relatively to $ u_j $. Therefore, it is admissible that \textit{all} Bethe roots organize in $ n $-strings, i.e., a set of $ n $ Bethe roots with the imaginary part separated by $ i $ (this is known as the \textit{string hypothesis} \cite{1971PThPh..46..401T}). Obviously, the same argument could be applied to the other wing associated to $ v $ rapidities. At finite temperature we should expect infinitely many strings of all possible sizes. The proper variables to describe these degrees of freedom are therefore densities of Bethe roots and holes \cite{1999tods.book.....T} associated to each of the strings and physical rapidities. The asymptotic Bethe equations ought to be written using these densities. 
We introduce the densities $ \rho_n $ and $ \bar{\rho}_n $ of Bethe roots and holes, reserving the index $ n=0 $ for densities of $ \theta 's $, $ n>0 $ for the densities of $ n $-strings of $ u's $, and $ n<0 $ for strings of $ v's $. 

The Bethe equations (\ref{eq:ABA1}),(\ref{eq:ABA2}),(\ref{eq:ABA3}) become \cite{Gromov:2008gj}
\beq
\varrho_n+\bar\varrho_n=\frac{1}{2}m \cosh(\pi\theta) \delta_{n0} -\sum_{r=-\infty}^{\infty} K_{n,r}*\varrho_r\;,
\eeq
where $*$ stands for the usual convolution
\beq
f*g=\int\limits_{-\infty}^{+\infty} d\theta' f(\theta-\theta')g(\theta')\;,
\eeq 
The kernels $K_{nm}$ are the derivative of the logarithm of the S-matrices between the several excitations ($u$ strings, $v$ strings and physical rapidities), see \cite{Gromov:2008gj} for details. For instance, 
\beq\label{eq:K0}
-K_{0,0}(\theta)\equiv K_0(\theta)=\frac{1}{2\pi i} \frac{d}{d\theta}
\log S_0^2(\theta)\;. 
\eeq
To find the ground-state at finite temperature ($ 1/L $) and infinite volume, we must minimize the free energy. The outcome is a set of equations constraining the densities, known as TBA equations\begin{equation}\label{eq:Ysystem}
\log Y_{n}+\delta_{n,0}mL\cosh\pi u=s\ast\log(1+Y_{n+1})(1+Y_{n-1})\,,
\end{equation}
where $ s=\frac{1}{2\cosh\pi u} $. These equations are written in terms of some thermodynamic functions $ Y_n $. In the ground state, they are simply related to the densities by $ \log Y_n={{\bar{\rho}_n}/{\rho_n}} $ for $ n\neq 0 $ and $ \log Y_0={{\rho_0}/{\bar{\rho}_0}} $. Moreover, the exact finite volume energy of the system is given by\begin{equation}\label{TBAE0}
E_0(L)=-\frac{m}{2}\int du\cosh(\pi u)\log(1+Y_{0})\,.
\end{equation}
Equivalently, if we assume  that $ \log{Y_n(u)} $ has no singularities in the physical strip $ |Im(u)|\leq\frac{1}{2} $, for any n, then the TBA equations are equivalent to the so called Y-system\begin{equation}
Y_{n}^{+}Y_{n}^{-}=(1+Y_{n+1})(1+Y_{n-1})\,, \la{Yststem}
\end{equation}
with the asymptotic "boundary conditions" $Y_{n}(u)\propto e^{-mL\delta_{n,0}\cosh\pi u}$ 
for large $ u $. Throughout this paper, we will often use the notation 
\beq
f^{\pm}=f(u\pm\frac{i}{2}) \,\,\, , \qquad f^{\pm \pm}=f(u\pm i)  \,\,\, , \qquad f^{[\pm n]}=f(u\pm \frac{in}{2})\,.
\eeq

\subsection{The Wick rotation method}
Using the TBA approach we obtained the value of the ground state energy of the system at finite volume. However, the solution involves an infinite number of coupled equations. The GKV method \cite{Gromov:2008gj} allows one to collapse these equations into a single integral equation.\footnote{Furthermore, it can be used to study excited states of the theory in a systematic fashion.} This method in strongly based on the Hirota equation 
\begin{equation}\label{eq:Hirota}
T_{n}^{+}T_{n}^{-}-T_{n-1}T_{n+1}=\Phi ^{[n]}\bar{\Phi}^{[-n]}\,,
\end{equation} 
which is related to the $ Y $-system (\ref{Yststem}) by the change of variables \begin{equation}\label{eq:Yfunc}
Y_{n}=\frac{T_{n+1}T_{n-1}}{\Phi ^{[n]}\bar{\Phi}^{[-n]}}\,.
\end{equation}
The $ Y $-system is actually a gauge invariant version of the Hirota equation, as can be seen by noting that it remains unchanged by the transformations\begin{eqnarray}\label{eq:GAUGE}
T_k & \rightarrow & g^{[k]}\bar{g}^{[-k]}T_k \nonumber \\
\Phi & \rightarrow & g^{+}g^{-}\Phi\nonumber \\
\bPhi & \rightarrow & \bar g^{+}\bar g^{-}\bPhi \,,
\end{eqnarray}
where $g$ and $\bar g$ are arbitrary functions. To ensure reality, $\bar g$ is the conjugate of $g$ and $\bar \Phi$ is the conjugate of $\Phi$.  

The crucial observation is that the general solution to Hirota equation (\ref{eq:Hirota}) is known! It involves a two by two determinant with
four unknown functions, see e.g. \cite{Zabrodin:1996vm}. Requiring a real solution we reduce this number to two functions (and their complex conjugates).
This means that effectively we only need to find these two functions to solve (\ref{eq:Hirota}) completely. Furthermore, as described above, there is a huge gauge symmetry of Hirota which relates different solutions to Hirota equation
yielding the same Y-functions. Using this gauge symmetry we can render one of the two remaining functions trivial \footnote{constant or a simple
polynomial, depending on the situation, see \cite{Gromov:2008gj} for details.}.
Thus, at the end of the day there is one function left unfixed. We can trade this function for another interesting unknown as we now
describe.
For the vacuum, the Y-functions for the two wings must be symmetric, $Y_k=Y_{-k}$. This does \textit{not} necessarily imply that $T_k=T_{-k}$
but it \textit{does} implies that $T_k$ and $T_{-k}$ are related by a gauge transformation $g(u)$. The knowledge of this function is
equivalent to the knowledge of the last unknown function in the general solution of Hirota. At the end of the day, the infinite set of integral equations for the $Y_n$ functions become a single integral equation for the gauge function $g$. It reads \cite{Gromov:2008gj}
\begin{equation}\label{eq:PCF}
g^2= -e^{i m L \sinh(\pi x)} \exp \left(2i\; \im\! \[K_0^-*\log\frac{\left(g^+\right)^2-1}{ | g^+ |^2-1} \] \right)\;.
\end{equation}
The ground-state energy can be expressed in terms of this gauge function $ g $, as\begin{equation}
E_0(L)=-\frac{m}{2}\int du\cosh(\pi u)\log\frac{\left(g^+\right)^2-1}{ | g^+ |^2-1}\frac{\left(\bar{g}^-\right)^2-1}{ | g^+ |^2-1}\,.  \la{energy}
\end{equation}
Therefore, the problem of computation of the exact ground-state energy at finite volume is now solved by a single equation.

This method can in principle be applied to many integrable theories. For instance, the finite volume ground-state of the model $ SU(2) $ Chiral Gross-Neveu \cite{Gross:1974jv} was solved by this method, resulting in the well-known Destri-de Vega (DdV) equation \cite{Destri:1992qk,Destri:1994bv}. This equation turns out to be exactly the same as (\ref{eq:PCF}) if we simply drop the denominator $ | g^+ |^2-1$ inside the logarithm to be left with 
\begin{equation}\label{eq:GN}
g^{2}=-e^{imL\sinh{(\pi x)}}\exp{\left(i\im K_0^{-} \ast\log{((g^+)^{2}-1)}\right)}\,.
\end{equation} 
The ground-state energy can again be expressed in terms of $ g $. Again, the result turns out to be exactly the same as in (\ref{energy}) once we drop the denominator $ | g^+ |^2-1$ in this expression.  In the last section, we will give a physical explanation for this distinction. 


\section{Lattice approach}
\subsection{Faddeev-Reshetikhin model}\label{FR}
An alternative method for the quantization of the $ SU(2) $ PCF was proposed in \cite{Faddeev:1985qu} by
Faddeev and Reshetikhin (FR). It is based on an integrable discretization of the model. In this approach the $SU(2)$ principal chiral field particles are the effective excitations around the anti-ferromagnetic vacuum of a quantum spin chain. 


The direct quantization by the quantum inverse scattering method \cite{Faddeev:1979gh,Takhtajan:1979iv,Sklyanin:1980ij} of the action (\ref{eq:action}) turns out to be problematic due to the structure of Poisson brackets. Let us recall why and review what was the approach of FR to circumvent this point. We introduce the currents $ L_{\mu}=\partial_{\mu}gg^{-1}=\sum_{a}L_{\mu}^{a}t^{a} $, where $ t^{a} $ are the generators of $ SU(2) $\footnote{These generators satisfy $ [t^{a},t^{b}]=i\epsilon^{abc}t^{c} $ and $ \tr{t^{a}t^{b}}=\delta^{ab} $}. By definition, they satisfy the flatness condition\begin{equation}\label{eq:currents}
\partial_{\mu}L_{\nu}^{a}-\partial_{\nu}L_{\mu}^{a}+i\epsilon^{abc}L_{\mu}^{b}L_{\nu}^{c}=0\,.
\end{equation}
We can rewrite the action (\ref{eq:action}) in terms of these currents, introducing a Lagrange multiplier field $ 2e_{0}^{2}\pi^{a}(x) $ to fix the constraint (\ref{eq:currents})\begin{equation}
S=-\frac{1}{2e_{0}^{2}}\sum_{a}\int dt\,dx\,\left((L_{0}^{a})^{2}+(L_{1}^{a})^{2}+2e_{0}^{2}\pi^{a}(\partial_{1}L_{0}^{a}-\partial_{0}L_{1}^{a}+i\epsilon^{abc}L_{1}^{b}L_{0}^{c})\right)\,.
\end{equation}
The current $L_0^{a}$ is not dynamical in the sense that it appears with no time derivatives. 
The equation of motion for this current reads \begin{equation}\label{eq:L0}
L_{0}^{a}=e_{0}^{2}i\epsilon^{amb}\pi^{m}L_{1}^{b}-e_{0}^{2}\partial_{1}\pi^{a}\,.
\end{equation}   
We can now compute the Poisson brackets of the current elements, such as $\lbrace L_{0}^{a}(x),L_{1}^{b}(y)\rbrace$. We simply need to notice that $\pi^a$ is the conjugated momentum to $L_{1}^{a}$ so that the equal time Poisson bracket between these two objects is simply $\{L_1^{a}(x),\pi^b(y)\}=\delta^{ab} \delta(x-y)$. Therefore\begin{equation}
\lbrace L_{0}^{a}(x),L_{1}^{b}(y)\rbrace =e_{0}^{2}i\epsilon^{abc}L_{1}^{c}(x)\delta(x-y)-e_{0}^{2}\delta^{ab}\delta'(x-y)\,.
\end{equation}
The derivative of the delta function term rules out the application of the quantum inverse scattering method \cite{Faddeev:1979gh,Takhtajan:1979iv,Sklyanin:1980ij}, see \cite{Dorey:2006mx,Maillet:1985ec,Maillet:1985ek,Maillet:1985fn} for further discussions of these non-local terms. 

To overcome the difficulties generated by this term, FR proposed two inspired steps \cite{Faddeev:1985qu} 
\begin{itemize}
\item First, the current Poisson brackets are modified. The derivative of delta function terms are dropped. 
\item Next, a new Hamiltonian is proposed. The new Hamiltonian yields the original equations of motion using the new Poisson brackets. 
\end{itemize}
Ultimately, the validity of these two seemingly dangerous steps is justified by the final results which match perfectly with all other approaches. The great advantage of the new Hamiltonian is that is allows for an integrable discretization. Physically, this Hamiltonian describes two interacting magnets. Once we discretize it, the diagonalization of the lattice model is equivalent to the one of an inhomogeneous $ SU(2) $ spin-s chain with $N$ sites ($ N $ is even) and inhomogeneity parameter $a$. These parameters need to be sent to infinity in a precise way which we review below. 

For any finite $s$, $N$ and $a$ the bare spin chain model is described by an exact set of Bethe equations. The idea of the lattice approach to the finite size spectrum of the $SU(2)$ principal chiral field is to use these exact equations and follow all the continuum limits carefully to derive all finite size corrections to the effective model. 

The Bethe equations for the inhomogeneous $ SU(2) $ spin-s chain with $N$ sites and $M$ Bethe roots \cite{Faddeev:1985qu} read 
\begin{equation}\label{eq:FRBAE}
\left(\frac{u_j+a+is}{u_j+a-is}\right)^\frac{N}{2}\left(\frac{u_j-a+is}{u_j-a-is}\right)^\frac{N}{2}=\prod_{k\neq{j}}^{M}\frac{{u_j-u_k+i}}{{u_j-u_k-i}}\,,
\end{equation}
where $ a $ plays the role of the inhomogeneity. These are the bare Bethe equations describing the magnonic excitations over the ferromagnetic vacuum. To make the bridge with the $ SU(2) $ PCF we need to consider the solutions corresponding to fluctuations around the anti-ferromagnetic vacuum.

In the large $N$ limit, the anti-ferromagnetic vacuum of this chain is composed by a Dirac sea of $N/2$ $2s$-strings of Bethe roots ($M=Ns$) \cite{A.:1982zz,Babujian:1983ae}. The claim is that the excitations around this vacuum are the $SU(2)$ PCF excitations. The limit is taken as follows
\begin{itemize}
\item The physical length of the model\footnote{This is the length of the space circle measured in unit of inverse infinite volume mass gap}, $mL$, is given by 
\beq
mL=2Ne^{-\pi a} \,.
\eeq
We should take $N,a\to \infty$ keeping this quantity fixed. In this limit, the holes in this vacuum become massive excitations with relativistic dispersion relation. 
\item The scattering matrix for these holes becomes the scattering matrix of the $SU(2)$ PCF when we take furthermore the large spin limit,\footnote{It would be very interesting to explore what the finite $s$ model would describe and what the connection between the lattice and the Wick rotation approaches in that case.} 
\beq
s\to \infty\,.
\eeq
\end{itemize}
As mentioned above, the physical rapidities of the $SU(2)$ PCF are identified with the hole excitations of the anti-ferromagnetic vacuum that we denote by $\tilde{u}^{2s}$ (see appendix \ref{AppA}),
\beq
\theta  \leftrightarrow  \tilde{u}^{2s} \,.
\eeq
To describe the physical excitations of the $SU(2)$ principal chiral field we also need the rapidities $u$ and $v$ introduced in (\ref{eq:ABA1})-(\ref{eq:ABA3}). These turn out to correspond to string excitations of sizes $2s+1$ and $2s-1$ respectively,  whose centers we denote by $u^{2s+1}$ and $u^{2s-1}$,
\beqa
u & \leftrightarrow & u^{2s+1}\\
v & \leftrightarrow & u^{2s-1}\,.
\eeqa
More precisely, the equations for $\tilde u^{2s}$, $u^{2s-1}$ and $u^{2s+1}$ can be derived from (\ref{eq:FRBAE}). For large length $Lm$ they coincide with (\ref{eq:ABA1})-(\ref{eq:ABA3}) provided we identity $\tilde u^{2s}$, $u^{2s-1}$ and $u^{2s+1}$ with $\theta$, $v$ and $u$, respectively (see also  \cite{1998solv.int.10007Z,Volin:2010cq})\footnote{Rigorously, this correspondence is valid when $ u $ and $ v $ are real.}. 
This is reviewed in appendix \ref{AppA} using the DdV approach. It is the first major evidence for the correctness of the FR discretization! We observe however that the FR model generates only a part of the Hilbert space of the $ SU(2) $ PCF. As showed in \cite{Faddeev:1985qu}, the number of holes is always even and only singlet states in one isotopic sector can be excited.
\begin{figure}[htb]
\centering
\includegraphics{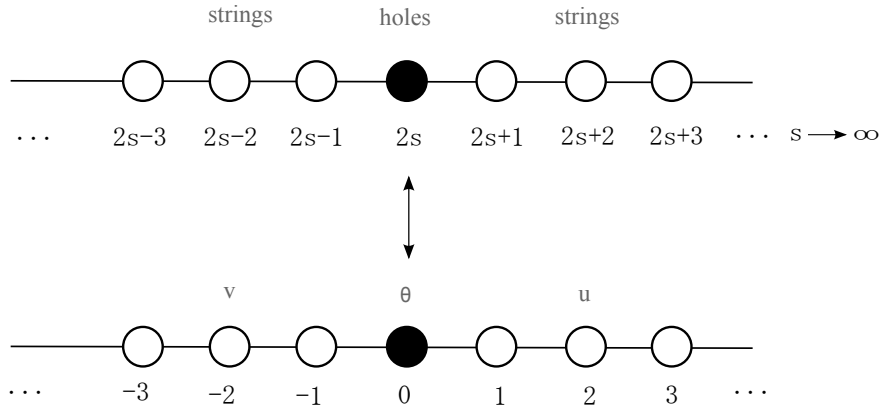}
\caption{In infinite volume, the degrees of freedom of the spin chain have a correspondence with those of the $ SU(2) $ PCF. However, a counting of the states of the spin chain shows that only singlets in one of the $ SU(2) $ sectors are generated}
\label{correspondence}
\end{figure}

\subsection{Finite Volume NLIE}
There are several techniques to study the finite size/temperature problem in quantum spin chains: the Destri-de Vega approach \cite{Destri:1992qk,Destri:1994bv}, the nonlinear integral equation (NLIE) method suggested in \cite{1990JPhA...23L.189K,1991JPhA...24.3111K}, the fusion procedure based \cite{Kulish:1981gi,Kulish:1980ii,Klumper,Kuniba:1993cn,Kuniba:1993nr}, and an hybrid formulation based on both fusion procedure and NLIE \cite{Suzuki:1998ve}. When the spins of the chain are in higher representations there is an important new feature of the anti-ferromagnetic vacuum: it is made out of a sea of \textit{bound-states} or \textit{Bethe strings}.\footnote{In fact, the Bethe roots condense into strings only in the large volume limit. Otherwise deviations to this behaviour appear.} This is the case for example for the FR model presented above. This characteristic of the models renders the hybrid formulation more suitable \cite{Suzuki:1998ve}; the DdV approach is not yet developed to comprise this case. 

The bare Bethe equations (\ref{eq:FRBAE}) follow from the diagonalization of a transfer matrix ${\bf{T}}_{2s}$ \cite{Faddeev:1985qu}. The \textit{exact} (bare) energy of the system is also given in terms of the transfer matrix. Interestingly, this transfer matrix is only one in a family of fused transfer matrices ${\bf{T}}_p$ with $p=1,2,\dots$. All these transfer matrices are related by simple functional relations. Together with the analytic properties of the transfer matrices, these functional relations have a unique solution. More precisely, the fusion procedure based method \cite{Kulish:1981gi,Kulish:1980ii,Klumper,Kuniba:1993cn,Kuniba:1993nr} converts these equations into integral equations of TBA type. Typically we end up with an infinite set of integral equations. Often these can be simplified into a finite set of equations \cite{Suzuki:1998ve}. 

In this section, we will start by reviewing how to obtain an infinite set of coupled equations of TBA type, using the fusion procedure. We will then derive a single non linear integral equation that allows to compute the finite volume ground-state energy of the $ SU(2) $ PCF. Our method is based on the proposals of \cite{Suzuki:1998ve,Hegedus:2003xd}. In \cite{Hegedus:2003xd}, it was obtained a set of two coupled integral equations. But, as we will see, this can be reduced to a single one equation.

The starting point of the fusion procedure is the introduction of a family of fused transfer matrices defined by
\begin{equation}\label{eq:T}
{\bf{T}}_{j}\equiv \sum_{l=1}^{j+1} \lambda_{l}^{(j)}\,,
\end{equation}
where \begin{equation}
\lambda_{l}^{(p)}\equiv  \prod_{m=l-1}^{p-1}\phi^{[2m-2s-p+1]}\prod_{n=0}^{l-2}\phi^{[2n+2s+1-p]}\frac{Q^{[p+1]}Q^{[-p-1]}}{Q^{[2l-p-1]}Q^{[2l-p-3]}}\,.
\end{equation}
In this expression, $ \phi(u)\equiv (u-a)^{\frac{N}{2}}(u+a)^{\frac{N}{2}} $ and $ Q(u)\equiv \prod_{j}({u-u_j}) $ is the Baxter polynomial. The Bethe equations (\ref{eq:FRBAE}) ensure that functions $ {\bf{T}}_{j} $ have no poles. Next we introduce another set of functions, the $Y$-functions. They are given by 
\begin{equation}\label{eq:Yj}
{\bf{Y}}_{j}\equiv \frac{{\bf{T}}_{j-1}{\bf{T}}_{j+1}}{\prod_{p=1}^{j}\phi^{[2p-j+2s]}\phi^{[-2p+j-2s]}}\,.
\end{equation}
From (\ref{eq:T}) the following alternative definition also follows:\begin{equation}\label{YT}
1+{\bf{Y}}_j(u)=\frac{{\bf{T}}_{j}^{+}{\bf{T}}_{j}^{-}}{\prod_{p=1}^{j}\phi^{[2p-j+2s]}\phi^{[-2p+j-2s]}}\,.
\end{equation}
\textit{By definition}, they obey the $Y$-system functional relations
\begin{equation}\label{eq:Ysys}
{\bf{Y}}_{j}^{+}{\bf{Y}}_{j}^{-}=(1+{\bf{Y}}_{j-1})(1+{\bf{Y}}_{j+1}) \,.
\end{equation} 
Recall that exactly the same equations arise in the TBA context, see (\ref{Yststem}). Of course, the derivation of the $Y$-system equations in both approaches is radically different.

Once supplemented with simple analytic properties of the $Y$-functions, the $ Y $-system equations can be converted into integral equations for ${\bf{Y}}_n$ which admit a unique solution! 
The analytic properties of the $ Y $-functions follow from those of ${\bf{T}}_j$, which in turn depend on Baxter polynomials $ Q $. In the large $ N $ limit, the zeroes of $Q$ are given in terms of $ 2s $-strings, and, for finite $ N $, we expect the deviation of the imaginary part of these zeroes to be small (less than $\frac{1}{4}$ according to \cite{deVega:1989pj}).

The determination of the zeroes of ${\bf{T}}_j$ in the ground state needs some care. The zeroes can appear in two different ways. First, there are zeroes coming from the factorization of some $ \phi $'s that are common in all terms of the expression for ${\bf{T}}_j$, (\ref{eq:T}). For clarity, we call these zeroes the \textit{trivial} zeroes. On the other hand, there are zeroes that appear due to the Baxter polynomials in the expression of ${\bf{T}}_j$.

The trivial zeroes appear only when $ j>2s $. When $ j\leq 2s $ there are no common $ \phi $'s in all terms of (\ref{eq:T}). So, if we let $ j=2s+k $ ($k>0$), then the trivial zeroes are the zeros of the common factor\begin{equation}\label{trivial}
\prod_{m=0}^{k-1}\phi^{[k-1-2m]}\,.
\end{equation}  

The remaining zeroes can be determined by considering the large $ N $ behaviour of ${\bf{T}}_j$. Such kind of analysis is nicely discussed in \cite{Volin:2010cq} which we will follow closely. For finite $ N $ we expect a small deviation from this behaviour as stated before.  In the large $ N $ limit, either the first or the last term of (\ref{eq:T}) is dominant, depending on whether the imaginary part of the argument is negative or positive, respectively. In this way, we can obtain the approximate positions of the complex zeroes. Explicitly,  if $\im u>0$,\begin{equation}
{\bf{T}}_{j}(u)\sim\frac{Q^{[-j-1]}}{Q^{[j-1]}}\prod_{n=0}^{j-1}\phi^{[2n+2s+1-j]}
\end{equation} 
and if $\im u<0$\begin{equation}
{\bf{T}}_{j}(u)\sim\frac{Q^{[j+1]}}{Q^{[-j+1]}}\prod_{n=0}^{j-1}\phi^{[-2n-2s-1+j]}\,.
\end{equation}

We now consider the ground-state configuration at large $ N $, where the Bethe roots organize in $ 2s $-strings. Therefore, for $j\leq{2s}$, some of the zeroes of the Baxter polynomial in the numerator cancel with the same zeroes of the Baxter polynomial in the denominator. Therefore, the position of the imaginary part of the zeros of ${\bf{T}}_{j}$ in the large $N$ limit is given by\begin{eqnarray}
\bullet &\,\, \pm\left({s+\frac{j}{2}-l}\right)\,\,\, \text{with}\,\,\, l=0,...,j-1,\,\,\,\, \text{for}\,\, j\leq 2s \label{Tl2s}\\ 
\bullet &\,\, \pm\left({s+\frac{j}{2}-l}\right)\,\,\, \text{with}\,\,\,  l=0,...,2s-1,\,\,\,\, \text{for}\,\, j> 2s  \label{Tg2s}\,.
\end{eqnarray}
With these considerations we found $ Nj $ zeroes of ${\bf{T}}_j$, for any $ j $. Indeed, for $ j\leq 2s $, all zeroes come from the Baxter polynomials. Taking into account (\ref{Tl2s}) and the fact that in ground-state we have $ \frac{N}{2}\,\, 2s$-strings, the number of complex zeroes  of  ${\bf{T}}_{j\leq 2s}$  is $ \frac{N}{2}\times 2j=Nj $. For $ j>2s $, we have $ N(j-2s) $ trivial zeroes according to (\ref{trivial}), and $ 2sN $ complex zeroes due to (\ref{Tl2s}). This gives again a total of $ Nj $ zeroes.

Moreover, a comparison of the number of zeroes in both sides of (\ref{eq:T}) shows that in the ground-state we do not expect extra zeroes on the real axis. Indeed, let $ n_h $ be the number of real zeroes of ${\bf{T}}_j$ excluding trivial zeroes, so that its total number is $ n_h+Nj $, for any $ j $. From (\ref{eq:T}), we see that the right hand side is the ratio of two polynomials of degree $ 2(j+1)Ns+Nj $ and $ 2(j+1)Ns $. Furthermore, due to Bethe equations (\ref{eq:FRBAE}) we know that the zeroes in the denominator are cancelled by the zeroes in the numerator. Therefore, the right hand side of  (\ref{eq:T}) is a polynomial of degree $ Nj $ due to the factors of $ \phi $ in each term $\lambda_{l}^{(j)}$. Equating the number of zeroes of both sides gives that $ n_h=0 $ and we conclude that in ground-state there are no zeroes on the real axis.

\begin{figure}[htb]
\begin{center}
\scalebox{0.56}{
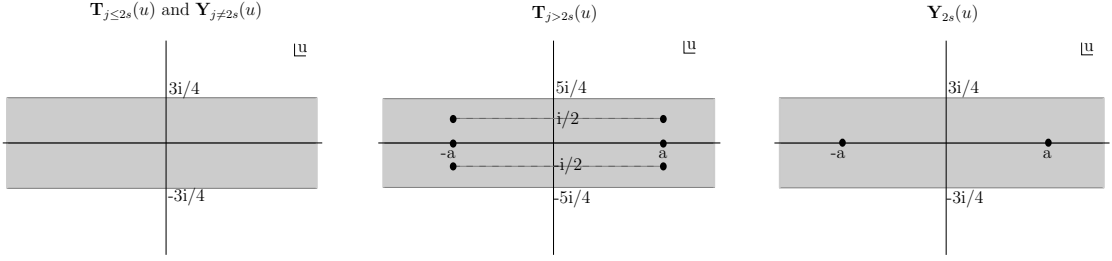}
\end{center}
\caption{The regions ANZC of the functions ${\bf{T}}_j$ and ${\bf{Y}}_j$ are represented in grey, the black dots represent $ \frac{N}{2} $-fold isolated zeroes and $ a $ is the inhomogeneity parameter.}
\label{regions}
\end{figure}

Using this information we can determine the region where ${\bf{T}}_j$ are analytic, non-zero with constant asymptotic behaviour (ANZC). The properties of ${\bf{Y}}_j$ can be extracted from those of ${\bf{T}}_j$ using the expression (\ref{eq:Yj}). The functions ${\bf{Y}}_j$ do not have zeroes nor poles in the strip $ |\im u|\leq\frac{3}{4} $, except for ${\bf{Y}}_{2s}$, which has $ \frac{N}{2} $-fold zeroes at the positions $ \pm a $. This follows from the observation that all trivial zeroes of ${\bf{T}}_j$ in the strip  $ |\im u|\leq\frac{3}{4} $ cancel with similar zeroes of the denominator in (\ref{eq:Yj}), except for the case ${\bf{Y}}_{2s}$. Recall that $a$ is going to be sent to infinity in a particular scaling limit but it is crucial to keep these zeroes at this point; as anticipated above they will be responsible for the dynamical mass generation of the effective excitations. All these ANZC regions are summarized in the figure \ref{regions}.

We will derive integral equations for ${\bf{Y}}_{j}$ from the $ Y $-system (\ref{eq:Ysys}). Let us define a new function ${\tilde{\bf{Y}}}_{2s}\equiv \frac{{\bf{Y}}_{2s}}{\phi}  $ which is now ANZC in the strip $|\im u|\leq\frac{3}{4}$. This function satisfies the equation\begin{equation}\label{ytilde}
{\tilde{{\bf {Y}}}}_{2s}^{+}{\tilde{{\bf{Y}}}}_{2s}^{-}=\frac{(1+{\bf{Y}}_{j-1})(1+{\bf{Y}}_{j+1})}{\phi^{+}\phi^{-}}\,.
\end{equation}
We now consider the shift operator $ s=\frac{1}{2\cosh\pi u} $ which has the following property: for an analytic and bounded function $ f(u) $ in the strip $ |\im u|\leq \frac{1}{2} $, we have\begin{equation}
s\ast[f^{+}+f^{-}]=f\,.
\end{equation}
We take the logarithm and apply the shift operator in both sides of  (\ref{eq:Ysys}) for $ j\neq 2s $ to get
\begin{equation}\label{NLIEY}
\left\{ \begin{array}{rl}\medskip
  \log{{\bf{Y}}_1}=&s\ast\log(1+{\bf{Y}}_{2})\\ 
  \log{{\bf{Y}}_j}=&s\ast\log(1+{\bf{Y}}_{j-1})(1+{\bf{Y}}_{j+1}),\;\;\;\; j\neq 2s\,. \end{array} \right.
\end{equation}
Similarly, doing the same to equation (\ref{ytilde}) we get\begin{equation}
\log{{\bf {Y}}_{2s}}=\log{\phi}-s\ast\log{\phi^{+}\phi^{-}}+s\ast(1+{\bf{Y}}_{j-1})(1+{\bf{Y}}_{j+1})\,.
\end{equation}
When we consider the continuum limit in which $a\to\infty$ and $ N\to\infty $, with the quantity $mL=2Ne^{-\pi a}$ held fixed, the terms involving $ \phi $'s become $-mL \cosh(\pi u)$. Finally the equation is written as\begin{equation}\label{y2s}
\log{{\bf {Y}}_{2s}}=-mL \cosh(\pi u)+s\ast(1+{\bf{Y}}_{j-1})(1+{\bf{Y}}_{j+1})\,.
\end{equation}

The ground-state energy can be expressed in terms of ${\bf{Y}}_{2s}$, by the following considerations. According to \cite{Faddeev:1985qu}, the energy of a state is given by\begin{equation}\label{en}
E=\frac{1}{i\delta}\log \frac{Q(-a+is)}{Q(-a-is)}\frac{Q(a-is)}{Q(a+is)} \,,
\end{equation}
where we have introduced the lattice spacing $ \delta $ to anticipate the continuum limit. Using the expression (\ref{eq:T}), it is possible to express the Baxter polynomials in terms of $ {\bf{T}}_{2s} $ and $ \phi $. The reason is that when ${\bf{T}}_{2s} $  is evaluated at $ \pm a \pm i/2 $ almost all terms vanish due to $ \phi(\pm a)=0 $. It remains a single one which involves precisely the Baxter polynomials appearing in (\ref{en}). Then, we get\begin{equation}\label{almosten}
E=\frac{1}{\delta}\log \frac{{\bf{T}}_{2s}^{+}(a)}{{\bf{T}}_{2s}^{+}(-a)}+ \frac{1}{\delta}\sum_{n=0}^{2s-1}\log \frac{\phi(-a+i(n+1))}{\phi(a+i(n+1))}\,.
\end{equation}
The term involving $ \phi $'s is present in all states and gives an infinite constant contribution when the continuum limit is taken. Therefore, we will ignore it. Focusing only on the first term of (\ref{almosten}) we can rewrite it using the previous analytic properties of $ {\bf{T}}_{2s} $ in the ground-state. In particular, with the equation (\ref{eq:Yj}) we can write\begin{equation}
{\bf{T}}_{2s}=s\ast \log(1+{\bf{Y}}_{2s})+  \text{terms involving $ \phi $'s}\,.
\end{equation}
Plugging this result in (\ref{almosten}) and ignoring once more the divergent terms involving $ \phi $'s, we finally obtain the ground-state energy in the continuum limit
\begin{equation}\label{eq:enspin}
E_0 =-\frac{m}{2}\int\limits_{-\infty}^{+\infty}du\cosh(\pi u)\log{(1+{\bf{Y}}_{2s}(u))}\,.
\end{equation}
As before, we took the continuum limit by letting $ \delta \to 0 $ and $ a\to\infty $ with the quantity $m L\equiv  2 N e^{-\pi a}  $ held fixed. Thus, we derived a system of equations resembling the TBA. Indeed, when $ s\to \infty $ the identification $ {\bf{Y}}_{2s+j} =Y_{j} $ leads to a match of both approaches.

We want now to truncate this system of equations using only functions which are natural from the lattice discretization point of view. The first step in this direction was given by Suzuki \cite{Suzuki:1998ve}, by the introduction of the following function\begin{equation}
a_{0}(u)\equiv\frac{\lambda_{1}^{(2s)}(u+\frac{i}{2})+...+\lambda_{2s}^{(2s)}(u+\frac{i}{2})}{\lambda_{2s+1}^{(2s)}(u+\frac{i}{2})}\,.
\end{equation}
This function (and its complex conjugated) allows to truncate the number of equations in the $ Y $-system as we will see.
It can be written in the following way\begin{equation}
a_0(u)={\bf{T}}_{2s-1}\frac{Q^{[2s+2]}}{Q^{[-2s]}}\frac{\phi}{\prod_{n=0}^{2s-1}\phi^{[2n+2]}}\label{a0}\,.
\end{equation}
A consequence of the definition is \begin{equation}\label{1a0}
1+a_{0}(u)=\frac{{\bf{T}}_{2s}^{+}}{\prod_{m=0}^{2s-1}\phi^{[2m+2]}}\frac{Q^{[2s]}}{Q^{[-2s]}}
\end{equation}
and also of great importance is the relation 
\begin{equation}\label{eq:trunc2}
(1+a_{0})(1+\bar{a}_{0})=1+{\bf{Y}}_{2s}\,,
\end{equation}
which is again a simple consequence of the definition of $a_0$ and of ${\bf{Y}}_{2s}$. The expression (\ref{eq:trunc2}) truncates the number of equations in the $ Y $-system. Thus, one can think of the functions $a_{0}$ (and its complex conjugate) as a re-summation of the contribution of the $ {\bf{Y}} $ functions with index larger than $ 2s-1 $. Moreover, we see that the ground-state energy can also be expressed in terms of $ a_0 $ as\begin{equation}\label{E_a0}
E_0 =-\frac{m}{2}\int\limits_{-\infty}^{+\infty}du\cosh(\pi u)\log{(1+a_0)(1+\bar{a}_0)}\,.
\end{equation}

\begin{figure}[htb]
\begin{center}
\scalebox{0.6}{
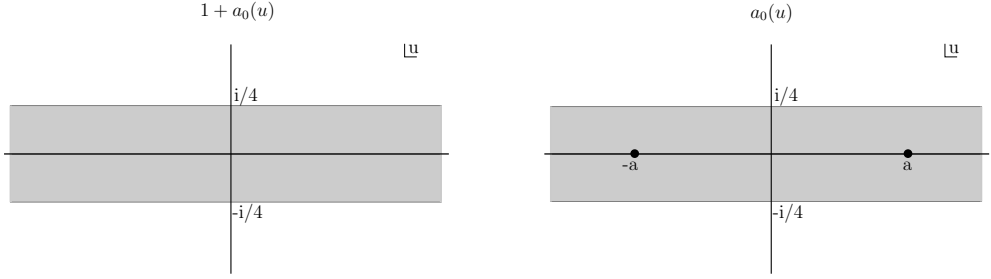}
\end{center}
\caption{The regions ANZC of the functions $ 1+a_{0}(u) $ and $ a_{0}(u) $ are represented in grey, the black dots represent $ \frac{N}{2} $-fold isolated zeroes and  $ a $ is the inhomogeneity parameter.}
\label{ANZCaux}
\end{figure}

From the definition, one can determine the ANZC regions of the functions $ a_0 $ and $ 1+a_0 $ which will be useful for what follows. They are summarized in the figure \ref{ANZCaux}.

With these analytic properties, we are in condition to perform the trick of \cite{Suzuki:1998ve}. Consider the function ${\bf{T}}_{2s}$ which is ANZC in the region $|\im u|\leq\frac{3}{4}$, then by Cauchy's theorem

\begin{figure}
\centering
\scalebox{0.8}{
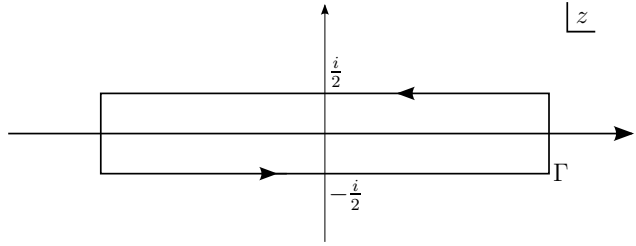}
\caption{The contour $ \Gamma $ encircles an ANZC region of the function ${\bf{T}}_{2s}(z)  $}
\label{contour}
\end{figure}

\begin{equation}\label{int2}
\oint_{\Gamma}dz\frac{d\log {\bf{T}}_{2s}(z)}{dz}e^{-ikz}=0\,,
\end{equation} 
where $\Gamma$ encircles the real axis counter-clockwise, in the upper plane with $z=\frac{i}{2}$ and in the lower plan with $z=-\frac{i}{2}$ (see figure \ref{contour}).

The relation (\ref{int2}) allows one to express the Baxter polynomials $ Q $ in terms of $1+a_0$ and $1+\bar{a}_0$ using the expression (\ref{1a0}). As explained in appendix \ref{AppB}, we obtain
\begin{eqnarray}\label{sysa0}
\left\{ \begin{array}{rl}\medskip
  \log{{\bf{Y}}_1}\,=&s\ast\log(1+{\bf{Y}}_{2})\\ \medskip
  \log{{\bf{Y}}_j}\,=&s\ast\log(1+{\bf{Y}}_{j-1})(1+{\bf{Y}}_{j+1}),\;\;\;\; j=1,...,2s-2\\ \medskip
  \log{{\bf{Y}}_{2s-1}}\,=&s\ast\log(1+{\bf{Y}}_{2s-2})+s\ast\log(1+a_0)(1+\bar{a}_0)\\ \medskip
  \log a_0\,=&-mL\cosh{\pi u}+\frac{1}{2}K_0\ast\log(1+a_0)-\frac{1}{2}K_0^{++}\ast\log(1+\bar{a}_0)-\frac{1}{2}\log(1+\bar{a}_0)\\
  &+s\ast\log(1+{\bf{Y}}_{2s-1})\,,
  \end{array} \right.
\end{eqnarray}
where we have taken as before the continuum limit when computing the terms involving the several $\phi$'s. Also, the convolutions with $ K_{0}^{++} $ are understood in the principal part sense. Note that at this point we did not take any limit on the spin $s$; it can be arbitrary so far.

As observed by Hegedus \cite{Hegedus:2003xd}, the limit $s\to\infty$ leads to an extra symmetry, due to the fact that the $ Y $-system becomes symmetric with respect to the massive node, ${\bf{Y}}_{2s}$, \begin{equation}
{\bf{Y}}_{2s+1}={\bf{Y}}_{2s-1}\,.
\end{equation}
This observation can be used to reduce the previous set of equations to a single equation for $ a_0 $. 
To derive this we observe that
\beqa
s\ast\log(1+{\bf{Y}}_{2s-1})&=&\frac{1}{2}s\ast \log(1+{\bf{Y}}_{2s+1}) (1+{\bf{Y}}_{2s-1})\label{secondline} \\
&=&\frac{1}{2}\left(\log {\bf{Y}}_{2s}+mL\cosh(\pi u)\right) \label{thirdline} \\
&=&\frac{1}{2}\left(\log((1+a_0)(1+\bar{a}_{0})-1) + m L \cosh(\pi u)\right)\,.\label{fourthline}
\eeqa
The first equality uses ${\bf{Y}}_{2s-1}={\bf{Y}}_{2s+1}$ which is the symmetry restored as $s\to \infty$ as described above. In the second equality, we used (\ref{y2s}). The last equality follows from (\ref{eq:trunc2}). 
%
Therefore the system of equations (\ref{sysa0}) can be simply rewritten as 
\begin{equation}\label{eq:a02}
\begin{split}
a_0\,=\,&e^{-\frac{mL}{2}\cosh{\pi u}}\exp \Bigl[ \frac{1}{2}K_0\ast\log(1+a_0)-\frac{1}{2}K_{0}^{++}\ast\log(1+\bar{a}_0)+\frac{1}{2}\log\Bigl((1+a_0)-(1+\bar{a}_0)^{-1}\Bigr) \Bigr]\,.
\end{split}
\end{equation}


\section{Equivalence of both approaches}
As reviewed in the previous sections, the Wick rotation approach and the lattice discretization approach to the finite size problem of the $SU(2)$ PCF lead to two seemingly distinct set of integral equations, (\ref{eq:g2}) and (\ref{intro}). In \cite{Gromov:2008gj,Beccaria:2010gq} numerical evidence was given in favour of their equivalence. The purpose of this section is to provide an analytic prove of their interchangeability. 

Comparing the formulas for the energy of the ground state given in both approaches, (\ref{eq:g2}) and (\ref{E_a0}), we are inspired to try 
\begin{equation}\label{eq:change}
1+a_0=\frac{(g^+)^2-1}{|g^+|^2-1} \,,
\end{equation}
so that the expressions for the ground state energy take the same functional form. 
Indeed, with this simple change of variables we can prove the equivalence of the two approaches. First let us re-write the GKV equation (\ref{eq:g2}) by shifting the contour of integration as
\begin{equation}\label{eq:A}
(g^+)^2=-e^{-mL\cosh{\pi u}}\exp\left(K_{0}\ast\log{\frac{(g^+)^2-1}{|g^+|^2-1}}-K_{0}^{++}\ast\log{\frac{(\bar{g}^{-})^2-1}{|g^+|^2-1}}-\log{\frac{(\bar{g}^{-})^2-1}{|g^+|^2-1}}\right)\,.
\end{equation} 
This is actually the form of the equation which is most suited for numerics \cite{Gromov:2008gj}. Convolutions with $K_0^{++}$ are understood as principal value integration. Under the identification (\ref{eq:change}), the equation (\ref{eq:a02}) coincides precisely with (\ref{eq:A})! 

We observe that the approach followed in \cite{Hegedus:2003xd}, resulting in equations (\ref{intro}), is equivalent to ours. Indeed, the  equations (\ref{intro}) can be reduced to a single one. According to \cite{Hegedus:2003xd}, the functions $ a $ and $ \bar{a} $ satisfy\begin{equation}
(1+a)(1+\bar{a})=1+{\bf{Y}}_{2s+1}\,.
\end{equation}
Using the trick described in (\ref{secondline}-\ref{fourthline}) we obtain precisely (\ref{eq:a02}).


\section{Conclusion and discussion}\label{conclusion}
We have shown analytically the equivalence between the approaches studied in this paper: the Wick rotation and the integrable discretization based methods. Indeed, the change of variable (\ref{eq:change}) leads to a perfect match between equations (\ref{eq:g2}) and (\ref{eq:a02}). This also confirms the numerical agreement presented in \cite{Gromov:2008gj,Beccaria:2010gq}.

The GKV equation for the $ SU(2) $ PCF (\ref{eq:PCF}) has a striking similarity with the typical DdV equation (\ref{eq:GN}), with an important difference: the denominator $ |g^+|^2-1 $ appearing in the convolutions of (\ref{eq:g2}) is absent in (\ref{eq:GN}). The DdV approach converts the bare Bethe equations of a lattice model into an exact integral equation for a counting function (see appendix \ref{AppA}). This integral equation can be used to compute the exact anti-ferromagnetic ground-state energy for any number of sites, $ N $.
When the continuum limit of the spin chain is considered, this approach gives the finite volume corrections of the corresponding field theory. Typically $ g^2 $ appearing in (\ref{eq:g2}) and (\ref{eq:GN}) can be identified with the counting function in the continuum limit. Unfortunately, this method is designed to study only lattice models with anti-ferromagnetic vacuum made of real Bethe roots.

When the vacuum is a sea of complex Bethe roots, the DdV technique is not yet developed to comprise this case. Generally, the study of spin chains at very large $ N $ is made using the string hypothesis. This essentially means that \textit{all} the complex Bethe roots organize themselves into sets with the same real part (the \textit{center} of the string) and imaginary parts separated by $ i $. When the number of sites is not very large, it is assumed that the corrections are exponentially suppressed in $ N $. 

In principle, we could fuse the bare FR Bethe equations (\ref{eq:FRBAE}) to obtain some equations for the centers of the strings. Since the centers are real, the DdV is suitable. More importantly, the corrections given by DdV are exponentially suppressed in $ mL \sim Ne^{-\pi a} \ll N  $, so we could expect that the string hypothesis would be justified. Such approach is done in appendix \ref{AppA}, but the result is similar to (\ref{eq:GN}), i.e., the GKV equation (\ref{eq:g2}) without the denominator $ |g^+|^2-1 $.

The explanation for this mismatch is that the deviation to the string hypothesis is not negligible, because the number of Bethe roots $ M=Ns $ also goes to infinite. This is a somehow subtle point: when we have $N\rightarrow \infty$ and a finite number of roots which form strings, these strings are extremely vertical. More precisely, the deviations from the string hypothesis are exponentially suppressed in $N \gg N e^{-\pi a}$. However, when the number of Bethe roots also scales with $N$ the suppression becomes much smaller and the string hypothesis cannot be used, see also \cite{deVega:1989pj}. In other words, we see that the denominators in the GKV equation  for the $SU(2)$ PCF  have a very nice physical meaning: they are a manifestation of the string hypothesis deviation in the vacuum of the underlying lattice model\footnote{In the cases where the DdV method is suitable, as in the $ SU(2) $ Chiral Gross-Neveu, the exact DdV equation is recovered by the Wick rotation approach. It is well known that the $ SU(2) $ Chiral Gross-Neveu model has a lattice formulation through the simple $XXX_{\frac{1}{2}}$ inhomogeneous spin chain (see for instance \cite{Volin:2010cq}). The anti-ferromagnetic vacuum of this spin chain is made of real Bethe roots and, consequently, the DdV can be directly applied to the bare Bethe equations.}.

Unfortunately, the FR model does not generate the whole Hilbert space of the $ SU(2) $ PCF \cite{Destri:1987zu,Faddeev:1985qu}: the number of holes is always even and only the singlets in one isotopic sector can be excited. Nevertheless, a modification of the NLIE (\ref{intro}) to include the one particle excited states was proposed and numerically tested with success \cite{Hegedus:2004xd,Beccaria:2010gq}. It would be interesting to check if it still agrees with the GKV approach which is claimed to be valid for all excited states. We also hope that the comparison between the two approaches made in this paper can help to determine a perfect discretization of the model, comprising all the excited states.

One of the main motivations for the study of the present paper came
from the study of the finite size spectrum of planar AdS/CFT
describing strings in $AdS_5\times S^5$ and $\mathcal{N}=4$
supersymmetric Yang-Mills theory. The asymptotic spectrum is given by
Beisert-Eden-Staudacher equations \cite{BS,BES}, the finite size
$Y$-system was proposed in \cite{Y} and the corresponding TBA equations
were presented in \cite{TBA1,TBA2,TBA3}. These three results are the
analogue of equations (\ref{eq:ABA1}-\ref{eq:ABA3}), (\ref{Yststem}) and (\ref{eq:Ysystem}) for the $SU(2)$ PCF. The analogue
of (\ref{eq:g2}), i.e., a finite number of equations describing the full planar
spectrum of AdS/CFT, is still not known despite many interesting
recent advances \cite{Bel,Heg,GKLT}. It would be interesting to also
try to look for these finite set of equations from the lattice
discretization point of view. Furthermore, if such DdV like equations
are found from the Wick rotation method it would be interesting to
understand their physical meaning. Will they contain denominator terms
like (\ref{eq:g2})? If so, do they signal some intricate structure of a lattice
vacuum as in the case of the $SU(2)$ PCF? As a warm up, and to gain
some more physical intuition about the form of the final set of
non-linear equations, we could generalize the lattice method to the
$SU(N)$ PCF which was considered using the Wick rotation method in \cite{Kazakov:2010kf,Volin:2010cq,1998solv.int.10007Z}.


\section*{Acknowledgements}
I would like to thank Miguel Costa, Dmytro Volin, João Penedones, Nikolay Gromov and Vladimir Kazakov for comments and useful discussions.  I would like to thank Pedro Vieira for suggesting this problem and for guidance. The research has been supported in part by the Province of Ontario
through ERA grant ER 06-02-293. Research at the Perimeter Institute is supported in part
by the Government of Canada through NSERC and by the Province of Ontario through MRI. This work was partially
funded by the research grants PTDC/FIS/099293/2008 and CERN/FP/109306/2009.


\appendix

\section{DdV applied to the fused FR Bethe equations}\label{AppA}
In this appendix, we will analyse the FR Bethe ansatz equations (\ref{eq:FRBAE}) both in infinite and finite size, using the Destri-de Vega approach.
First, we want to study the large volume solutions of the FR Bethe ansatz equations (\ref{eq:FRBAE}) corresponding to fluctuations around the anti-ferromagnetic vacuum. We expect these effective excitations to have the same dynamics as the excitations of the $ SU(2) $ PCF. To accomplish this, we will use the DdV approach. This technique is often applied to the computation of finite volume spectrum of quantum field theories with a discretization admitting an anti-ferromagnetic vacuum  made of real Bethe roots. As we saw, our spin chain is a sea of complex Bethe roots which condense into $ 2s $-strings, when the number of sites $ N $ is very large. In this limit, we can then fuse the bare Bethe equations, which amounts to multiplying the equations associated to the same string. In the end, we will get equations for the centers of the strings. We will apply the DdV technique to these fused equations, and ignore the finite volume corrections. In the continuum limit, we expect therefore to obtain the large volume description of the $ SU(2) $ PCF, i.e., the Asymptotic Bethe ansatz equations.

The finite size study of these equations by the DdV approach has an instructive premeditation. In the string hypothesis, it is assumed that the corrections to it are exponentially suppressed in $ N $. As we will see below, the DdV approach applied to the fused FR bare Bethe equations, gives, in the continuum limit, corrections exponentially suppressed in $ \sim Ne^{-\pi a} \ll N $. Therefore, a priori, we could expect that the string hypothesis deviation should be subleading compared to the DdV finite volume corrections. However, it will turn out that the string hypothesis deviation is not negligible, because the number of Bethe roots $ M=Ns $ also goes to infinity. In other words, the string hypothesis is no longer exponentially precise in $N$ when the number of strings also grows with $N$. The effect of the curvature of the strings will be then manifest, when we compare the result with the correct GKV equation (\ref{eq:PCF}).

In the large $ N $ limit, the fused FR Bethe ansatz equations (\ref{eq:FRBAE}) are obtained by multiplying the Bethe roots belonging to the same string. After taking the logarithm, we obtain\begin{equation}\label{TFRBAE}
\frac{N}{2}\left(\theta_{n,2s}(u_{j}^{n}-a)+\theta_{n,2s}(u_{j}^{n}+a)\right)=2\pi Q_{j}^{n}+\sum_{m=1}^{\infty}\sum_{k=1}^{\xi_{m}}\Xi_{n,m}(u_{j}^{n}-u_{k}^{m})\,,
\end{equation}
where $ u_{k}^{m} $ is a center of a string of size $ m $, $\xi_{m}$ denotes the number of strings of length $ m $ and $ Q_j^{n} $ is an integer or half-integer (depending on the particular configuration considered). 

The functions $ \theta_{n,m} $ and $ \Xi_{n,m} $ have the following expressions\begin{equation}
\theta_{n,m}(u)=-\dashint d\omega \frac{ie^{i\omega u}}{2 \omega \sinh\left(\frac{|\omega|}{2}\right)}\left(e^{-\frac{|n-m|}{2}|\omega|}-e^{-\frac{(n+m)}{2}|\omega|}\right)
\end{equation}\begin{equation}
\Xi_{n,m}(u)=-\dashint d\omega e^{i\omega u}\Bigl[ \frac{i}{\omega}\coth\left(\frac{|\omega|}{2}\right)\left(e^{-\frac{|n-m|}{2}|\omega|}-e^{-\frac{(n+m)}{2}|\omega|}\right)+\frac{i}{\omega}\delta_{n,m} \Bigr]\,.
\end{equation}
The integrals are understood in the principal part sense, due to the singularity at the origin.

For simplicity, we will consider a state with $ M $ strings of size $ 2s $, $ N_R $ strings of size $ 2s+1 $ and $ N_L $ strings of size $ 2s-1 $, with $ Q_j^{n} $ being integer. Anticipating a result, we will denote $ u_{k}^{2s+1} $ by $ u_{k} $, $ u_{k}^{2s-1} $ by $ v_k $ and the holes $ \tilde{u}_{k}^{2s} $ by $ \theta_{k} $. We recall that holes correspond to the solutions of equation (\ref{TFRBAE}), distinct from the Bethe roots.
We introduce the simpler notation
\begin{eqnarray}
\mathcal{P}_{h}(u)&=&\theta_{2s,2s}(u-a)+\theta_{2s,2s}(u+a)\label{ker1}\nonumber\\
\mathcal{P}_{u}(u)&=&\theta_{2s+1,2s}(u-a)+\theta_{2s+1,2s}(u+a)\nonumber\\
\mathcal{P}_{v}(u)&=&\theta_{2s-1,2s}(u-a)+\theta_{2s-1,2s}(u+a)\nonumber\\
\mathcal{H}(u)&=&\Xi_{2s,2s}(u)\nonumber\\
\mathcal{U}(u)&=&\Xi_{2s,2s+1}(u)=\Xi_{2s+1,2s}(u)\nonumber\\
\mathcal{U}_u(u)&=&\Xi_{2s+1,2s+1}(u)\nonumber\\
\mathcal{U}_v(u)&=&\Xi_{2s+1,2s-1}(u)\nonumber\\
\mathcal{V}(u)&=&\Xi_{2s,2s-1}(u)\nonumber\label{ker2}\,.
\end{eqnarray}
Following the standard DdV approach, we define the \textit{counting function} $ Z(u) $ as\begin{equation}
Z(u)=\,\frac{N}{2}\mathcal{P}_{h}(u)-\sum_{k=1}^{M}\mathcal{H}(u-u_{k}^{2s})-\sum_{k=1}^{N_R}\mathcal{U}(u-u_{k})-\sum_{k=1}^{ N_L}\mathcal{V}(u-v_{k})\,.
\end{equation}

\begin{figure}
\centering
\scalebox{0.8}{
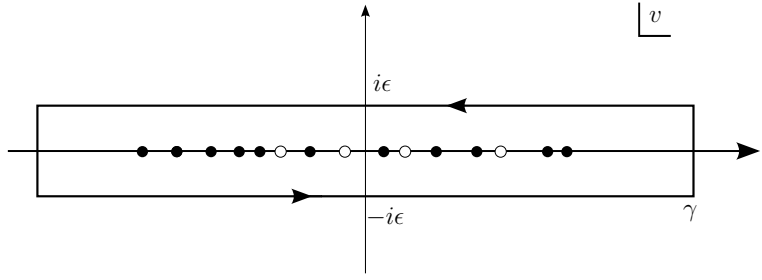}
\caption{The DdV contour $ \gamma $ encircles an analytic region (with non divergent asymptotic behaviour) of the function $ \Xi_{2s,2s}(v) $. The black dots represent the centers of $ 2s $-strings and the white dots represent the holes.}
\label{contourDDV}
\end{figure} 

The counting function has the property of $ Z(u_{k})=2\pi Q_k^{2s} $, where $u_{k}$ is a center of a $ 2s $-string or a hole. Essentially, the DdV technique allows to convert a sum over the centers of $ 2s $-strings into a sum over the holes in an exact way. In other words, using it, we can obtain the effective Bethe equations describing the excitations around the anti-ferromagnetic vacuum. Hence, it is an appropriate way of getting information about the dynamics of the physical particles.
Consider the identity\begin{equation}\label{sums}
\sum_{roots}\mathcal{H}(u-u_{k}^{2s})+\sum_{holes}\mathcal{H}(u-\theta_{k})=\oint_{\gamma}\frac{dz}{2\pi i}\mathcal{H}(u-z)\frac{d\log(1-e^{iZ(z)})}{dv}\,.
\end{equation}
where the contour $ \gamma $ is depicted in figure \ref{contourDDV}. Notice that the contour must not encircle poles of the function $ \mathcal{H}(u) $, and therefore $ 0<\epsilon <1 $.
Then, we rewrite this contour integral as\begin{equation}
\begin{split}
\int_{-\infty}^{\infty}\frac{dv}{2\pi}\mathcal{H}'(u&-v)Z(v) +\int_{-\infty}^{\infty}\frac{dv}{2\pi i}\mathcal{H}'(u-v+i\epsilon)\log(e^{-iZ(v-i\epsilon)}-1)\\
&-\int_{-\infty}^{\infty}\frac{dv}{2\pi i}\mathcal{H}'(u-v-i\epsilon)\log(1-e^{iZ(v+i\epsilon)})\,.
\end{split}
\end{equation}
We now use this result to substitute the sum over the Bethe roots in (\ref{sums}) by a sum over the holes, obtaining\begin{equation}
\begin{split}
(1+\frac{1}{2\pi}\mathcal{H}')\ast Z(u)&=\frac{N}{2}\mathcal{P}_{h}(u)+\sum_{k=1}^{\#holes}\mathcal{H}(u-\theta_{k})-\sum_{k=1}^{N_R}\mathcal{U}(u-u_{k})-\sum_{k=1}^{ N_L}\mathcal{V}(u-v_{k})\\
&-\frac{i}{2\pi}\mathcal{H}'^{[-2\epsilon]}\ast\log(1-e^{iZ(v+i\epsilon)})+\frac{i}{2\pi}\mathcal{H}'^{[2\epsilon]}\ast\log(e^{-iZ(v-i\epsilon)}-1)\,.
\end{split}
\end{equation}
Finally, after inverting the kernel $ 1+\frac{1}{2\pi}\mathcal{H}' $, we obtain the final DdV equation\begin{equation}
\begin{split}\label{eq:Zfinal} 
Z(u)=\,&\frac{N}{2}\mathcal{P}_{eff}(u)+\sum_{k=1}^{\#holes}\mathcal{H}_{eff}(u-\theta_{k})-\sum_{k=1}^{N_L}\mathcal{V}_{eff}(u-v_{k})-\sum_{k=1}^{N_R}\mathcal{U}_{eff}(u-u_{k})\\
&+iK_{eff}^{[2\epsilon]}\ast\log(e^{-iZ(v-i\epsilon)}-1)-iK_{eff}^{[-2\epsilon]}\ast\log(1-e^{iZ(v+i\epsilon)})\,,
\end{split}
\end{equation}
where\begin{eqnarray}
\mathcal{P}_{eff}(u)&=&\mathcal{F}^{-1}\left(\frac{\mathcal{F}(\mathcal{P}_{h}(u))}{1+\mathcal{F}(\frac{1}{2\pi}\mathcal{H}'(u))}\right)\label{eq:ddv1}\\
K_{eff}(u)&=&\mathcal{F}^{-1}\left(\frac{\mathcal{F}(\frac{1}{2\pi}\mathcal{H}'(u))}{1+\mathcal{F}(\frac{1}{2\pi}\mathcal{H}'(u))}\right)\label{k0}\\
\mathcal{V}_{eff}(u)&=&\mathcal{F}^{-1}\left(\frac{\mathcal{F}(\mathcal{V}(u))}{1+\mathcal{F}(\frac{1}{2\pi}\mathcal{H}'(u))}\right)\label{k1}\\
\mathcal{U}_{eff}(u)&=&\mathcal{F}^{-1}\left(\frac{\mathcal{F}(\mathcal{U}(u))}{1+\mathcal{F}(\frac{1}{2\pi}\mathcal{H}'(u))}\right)\label{k2}\\
\mathcal{H}_{eff}&=&\mathcal{F}^{-1}\left(\frac{\mathcal{F}(\mathcal{H}(u))}{1+\mathcal{F}(\frac{1}{2\pi}\mathcal{H}'(u))}\right)\label{k3}\,,
\end{eqnarray}
where $\mathcal{F}$ and $\mathcal{F}^{-1}$ denote the Fourier transform and inverse, respectively. The integrals in the Fourier transforms are again understood in the principal part sense.

The convolutions appearing in this equation give the finite size corrections. For the moment, we will drop them, and study the infinite size limit of this equation. In the continuum limit, $ N\to\infty $ and $ a\to\infty $, we define \begin{equation}
mL=2Ne^{-\pi a}
\end{equation}
and therefore the source term in equation (\ref{eq:ddv1}) can be rewritten as\begin{equation}
mL \sinh{\pi u}\,.
\end{equation}
After taking the limit $ s\to\infty $ in the kernels (\ref{k0}-\ref{k3}) and exponentiating, we obtain the following result for very large $ mL$\begin{equation}
e^{-imL\sinh\pi\theta_{j}}=\prod_{k\neq j}^{\# holes}S_{0}^{2}(\theta_{j}-\theta_{k})\prod_{k}^{N_L}\left(\frac{\theta{j}-v_{k}+\frac{i}{2}}{\theta_{j}-v_{k}-\frac{i}{2}}\right)\prod_{k}^{N_R}\left(\frac{\theta_{j}-u_{k}+\frac{i}{2}}{\theta_{j}-u_{k}-\frac{i}{2}}\right)\,.
\end{equation}
The auxiliary Asymptotic Bethe equations can be obtained similarly. Specifying (\ref{TFRBAE}) for   $n= 2s + 1 $, we have\begin{equation}\label{eq:strings}
\begin{split}
\frac{N}{2}\mathcal{P}_{u}(u_{k})=\,&2\pi Q_{j}^{2s+1}+\sum_{k=1}^{M}\mathcal{U}(u_{j}-u_{k}^{2s})\\
&+\sum_{k=1}^{N_R}\mathcal{U}_{u}(u_{j}-u_{k})+\sum_{k=1}^{ N_L}\mathcal{U}_{v}(u_{j}-v_{k})\,.
\end{split}
\end{equation}
The sum over the $2s$-strings can be converted in a sum over the holes, using DdV (ignoring the finite size corrections)\begin{equation}
\sum_{k=1}^{M}\mathcal{U}(u-u_{k}^{2s})=-\sum_{k=1}^{\# holes}\mathcal{U}(u-\theta_{k})+\frac{1}{2\pi}\mathcal{U}'\ast Z(u)\,.
\end{equation}
Using the expression (\ref{eq:Zfinal}), we obtain after exponentiation\begin{equation}
\prod_{k}\left(\frac{u_{j}-\theta_{k}-\frac{i}{2}}{u_{j}-\theta_{k}+\frac{i}{2}}\right)=\prod_{k\neq j}^{N_R}\left(\frac{u_{j}-u_{k}-i}{u_{j}-u_{k}+i}\right)\,.
\end{equation}
A similar equation could be derived for the strings of size $ 2s-1 $. Therefore we have obtained the Asymptotic Bethe ansatz equations for the $ SU(2) $ PCF.

We now turn to the second goal of this appendix, and consider the finite size corrections. We will restrict ourselves to the ground-state, in order to compare with the ground-state GKV equation (\ref{eq:PCF}). Hence, in the continuum limit and $ s\to\infty $, equation (\ref{eq:Zfinal}) can be rewritten as\begin{equation}
g^2=-e^{i mL\sinh\pi u}\exp\left(2i\,\im \[ K_0^{-}\ast\log((g^{+})^2-1)\]\right)\,.
\end{equation}
where $ g^{2}=e^{iZ(u)} $, and the shift in the contour of integration, $\epsilon$, was taken to be $ \frac{1}{2} $.
This equation resembles the GKV equation (\ref{eq:PCF}), with the significant difference that denominator $ |g^+|^2 -1 $ is absent here. We assumed that the corrections to this hypothesis should be exponentially suppressed in $ N $, and hence, negligible. But, as the number of Bethe roots is $ M=Ns $, and also goes to infinity, the corrections are significant. Our assumption is then manifestly wrong. Therefore, this extra denominator can be interpreted as the string hypothesis deviation in the anti-ferromagnetic vacuum of the associated discretization (see section \ref{conclusion}).

\section{Details of the derivation of the lattice NLIE}\label{AppB}
In this appendix, we give further details on the derivation of the hybrid equations (\ref{sysa0}). It is convenient to introduce the notation\begin{equation}
\hat{dl}f\equiv \int{du\frac{d \log f(u)}{du}e^{-iku}}\,.
\end{equation}
With this notation, the contour integral is rewritten as\begin{equation}\label{cont}
\hat{dl}{\bf{T}}_{2s}^{-}e^{-\frac{k}{2}}-\hat{dl}{\bf{T}}_{2s}^{+}e^{\frac{k}{2}}=0\,.
\end{equation}
Using (\ref{1a0}) and (\ref{cont}) we get\begin{equation}
\hat{dl}Q^{[\pm 2s]}=\frac{\Theta(\pm k)}{2\cosh \frac{k}{2}}\Bigl(\pm \hat{dl}(1+a_0)e^{\frac{k}{2}} \mp \hat{dl}(1+\bar{a}_0)e^{-\frac{k}{2}}+e^{\frac{\pm k}{2}}\sum_{m=0}^{2s-1}{\hat{dl}\phi^{[\pm 2m\pm 2]}}\Bigr)\label{q2s}\,,
\end{equation}where $ \Theta $ is the Heaviside step function. Next, we introduce a function $ \tilde{a}_0\equiv \frac{a_0}{\phi} $ which is ANZC in the region $ |\im u| \leq \frac{1}{4} $. Using (\ref{a0}) we have\begin{equation}\label{tildea0}
\hat{dl}\tilde{a}_0=\hat{dl}{\bf{T}}_{2s-1}+\hat{dl}Q^{[2s]} e^{-k}-\hat{dl}Q^{[-2s]}-\sum_{n=0}^{2s-1}\hat{dl}\phi^{[2n+2]}\,.
\end{equation}
From (\ref{YT}) with $ j=2s-1 $, we express $ {\bf{T}}_{2s-1} $ in terms of $ {\bf{Y}}_{2s-1} $\begin{equation}\label{T2s-1}
\hat{dl}{\bf{T}}_{2s-1}=\frac{1}{2\cosh \frac{k}{2}}\hat{dl}(1+{\bf{Y}}_{2s-1})+\frac{1}{2\cosh \frac{k}{2}} \sum_{p=1}^{2s-1}\hat{dl}({\phi^{[2p-j+2s]}\phi^{[-2p+j-2s]}})\,.
\end{equation} We then plug (\ref{q2s}) and (\ref{T2s-1}) in (\ref{tildea0}).  Finally, we invert the Fourier transform and integrate in $ u $ to get (\ref{sysa0}).


\end{document}